\documentclass{ifacconf}
\usepackage{amsmath,amssymb,amsfonts}
\usepackage{graphicx}      
\usepackage{natbib}        
\usepackage{enumitem}
\usepackage{dirtytalk}
\usepackage{rotating}
\usepackage{color}

\usepackage{algorithm}
\usepackage{algorithmic}
\usepackage{mathtools}
\mathtoolsset{showonlyrefs}
\newtheorem{theorem}{Theorem}[section]
\newtheorem{definition}{Definition}[section]

\newtheorem{assumption}{Assumption}[section]
\newtheorem{problem}{Problem}
\newcommand{\Dc}{\mathcal{D}}
\begin{document}
\begin{frontmatter}
\title{Stealthy Sensor Attacks Against Direct Data-Driven Controllers \thanksref{footnoteinfo}}
\thanks[footnoteinfo]{This work was supported by the Swedish Research Council grant 2024-00185.}
\author[Penn,KTH]{Sribalaji C. Anand} 
\address[Penn]{Department of Electrical and Systems Engineering, University of Pennsylvania, United States.}
\address[KTH]{School of Electrical Engineering and Computer Science, KTH royal institute of technology, Sweden. (e-mail: srca@kth.se)}
\begin{abstract}
This paper investigates the vulnerability of discrete-time linear time-invariant systems to stealthy sensor attacks during the learning phase. In particular, we demonstrate that a {data-driven} adversary, without access to the system model, can inject attacks that mislead the operator into learning an {unstable} state-feedback controller. We further analyze attacks that degrade the performance of data-driven ${H}_2$ controllers, while ensuring that the operator can always compute a feasible controller. Potential mitigation strategies are also discussed. Numerical examples illustrate the effectiveness of the proposed attacks and underscore the importance of accounting for adversarial manipulations in data-driven controller design.
\end{abstract}
\end{frontmatter}
%
\section{Introduction}
Data-driven control has gained significant attention in the literature due to its simplicity and effectiveness \citep{de2019formulas,van2023informativity}. This paradigm enables efficient controller design without requiring explicit identification of the underlying state-space model. In this paper, we study the resilience of data-driven control algorithms in the presence of adversarial attacks.

We consider a discrete-time (DT) linear time-invariant (LTI) plant, where sensor data are transmitted over a wireless network to a control center. The control center computes optimal control commands, used for reference tracking or set-point regulation, and sends them back to the plant. Since the controller does not know the plant model, it relies on a data-driven design method. An adversary is assumed to compromise the sensor measurements transmitted from the plant. Under this setup, we investigate the following problem:
\begin{problem}\label{prob:main}
Can a malicious data-driven adversary corrupt the sensor data so that the control center learns a sub-optimal feedback policy? \hfill $\triangleleft$
\end{problem}

By studying Problem~\ref{prob:main}, we make the following contributions to the literature:
\begin{enumerate}
    \item For data-driven stabilization, we propose a stealthy attack policy that misleads the operator into learning an {unstable} controller.
    \item Under our proposed attack, the operator can always compute a feasible controller. Such feasibility guarantee has not been addressed in the literature.
    \item For data-driven ${H}_2$-optimal control design, we formulate an optimization problem whose solution induces controller sub-optimality. Since this problem is non-convex, we propose a computationally efficient sub-optimal attack design algorithm.
    \item We also propose a modified attack policy that ensures the operator can always compute a feasible controller from the attacked dataset.
    \item We demonstrate the effectiveness of the proposed methods through numerical examples. In particular, we show that even a few iterations of our sub-optimal attack design algorithm can severely degrade the performance of the ${H}_2$ controller in closed-loop.
\end{enumerate}

These results position our work among the few papers that analyze cyber-physical attacks on data-driven control systems during the {learning} phase. A summary of related work is provided in Table~\ref{tab:summary}. Unlike many existing works, which primarily study bounded perturbation attacks, we explicitly design {stealthy} attacks. Moreover, we develop two different attack strategies targeting (i) data-driven stabilizing controller design and (ii) data-driven ${H}_2$-optimal control design. Another distinction is that most existing works assume that the adversary can inject attacks into both sensors and actuators, whereas we assume sensor-only attacks and still show that the adversary can significantly degrade operator performance. Finally, we emphasize that our framework assumes that both the operator and the adversary have no access to the true process dynamics; they rely solely on data.

\begin{table*}[t]
\centering
\caption{Summary of related works by system setting, attack type, studied problem, and related references. 
DT: discrete-time; CT: continuous-time; OM: output measurements; SM: state measurements; 
CD: clean data; ND: noisy data; NL: nonlinear; DoS: Denial-of-Service; VRFT: Virtual Reference Feedback Tuning. 
Data injection attacks target both controller and actuator channels.}
\label{tab:summary}
\begin{tabular}{|p{2.4cm}|p{2.2cm}|p{3cm}|p{5.3cm}|p{2.6cm}|}
\hline
 \textbf{Paper} & \textbf{System setting} & \textbf{Attack Type} & \textbf{Problem Studied} & \textbf{Related Work} \\ \hline
 \cite{zhao2023sparse} & DT; OM; ND & Sparse actuator attack & Attack detection against $\chi^2$ detector & \cite{zhao2022data} \\ \hline
 \cite{hu2025data} & CT; SM; CD & DoS attack & Data-driven stability criteria & \cite{liu2022resilient} \\ \hline
 \cite{li2024optimal} & DT; OM; ND & Data injection attack & Stealthy data-driven attack design & \cite{liu2024data} \\ \hline
\cite{shinohara2025detection} & DT; OM;  & Sparse sensor attack & Attack detection and identification under sparse observability & \cite{yan2025secure} \\ \hline
 \cite{sasahara2023adversarial} & DT; SM; & Data injection attack & Gradient-based bounded perturbation design against data driven LQ control & \cite{yu2022online} \\ \hline
 \cite{fotiadis2025deception} & CT; SM; & Eavesdropping attack & Optimal deception design against adversary &  \cite{mao2022decentralized}\\ \hline
\cite{yang2025data} & DT; OM; NL & Sensor attacks & Stabilizing control design against stealthy attacks & \cite{yin2025data,liu2024robust} \\ \hline
 \cite{krishnan2020data} & DT; OM; CD & Data injection attack  & Attack detection and identification &\cite{anand2023data} \\ \hline
 \cite{russo2021poisoning} &  DT; SM; & Data injection attack & Gradient-based bounded perturbation design against data driven VRFT control & \cite{russo2021data} \\ \hline
 \cite{lian2021online} & CT; SM; CD & Data injection attack & Attack resilient reinforcement learning & \cite{gao2022resilient} \\ \hline
 This work & DT; SM; CD & Sensor attacks & Data-driven stealthy destabilizing attacks, and attacks against $H_2$ control & \; \\ \hline
\end{tabular}
\end{table*}

\emph{Outline:} The remainder of this paper is organized as follows. Section~\ref{sec:problem} formulates the problem. Section~\ref{sec:design} presents an attack policy against data-driven stabilizing controller design. Section~\ref{sec:design2} discusses the attack policy against $H_2$-optimal controller design. Numerical examples illustrating the effectiveness of the proposed attacks are provided throughout the paper. Concluding remarks and possible future works are given in Section~\ref{sec:con}. 

\emph{Notation:} In this paper, $\mathbb{R}, \mathbb{C}$, and $\mathbb{Z}$ denote the sets of real numbers, complex numbers, and integers, respectively. An identity matrix of size $m \times m$ is denoted by $I_m$. A zero matrix of size $n \times m$ is denoted by $0_{n \times m}$. Let $x: \mathbb{Z} \to \mathbb{R}^n$ be a discrete-time signal with $x[k]$ denoting the value of the signal at time~$k$. The Hankel matrix associated with $x$ is denoted
\begin{equation}\label{Hankel_x}
X_{i,t,N} =
\begin{bmatrix}
x[i] & x[i+1] & \dots & x[i+N-1]\\
x[i+1] & x[i+2] & \dots & x[i+N]\\   
\vdots & \vdots & \ddots & \vdots\\
x[i+t-1] & x[i+t] & \dots & x[i+t-1+N]
\end{bmatrix},
\end{equation}
where the first subscript denotes the starting index of the signal, the second denotes the number of samples per column, and the last denotes the number of columns. When $t = 1$, the Hankel matrix is written as $X_{i,N}$. The notation $x_{[0,T-1]}$ denotes the vectorized segment of the signal from time $0$ to $T-1$, i.e.,
\[
x_{[0,T-1]} = \begin{bmatrix} x[0] & x[1] & \dots & x[T-1] \end{bmatrix}.
\]
A signal $x_{[0,T-1]}$ is said to be persistently exciting of order $L$ if the matrix $X_{0,L,T-L+1}$ has full rank $Ln$. The $2$-norm of a vector $x \in \mathbb{R}^n$ is denoted $\|x\|_2 = \sqrt{x^\top x}$. Let $A \in \mathbb{R}^{n \times n}$, then $\lambda_i(A), i \in \{1,\dots,n\}$ denotes the eigenvalues of $A$. A positive (semi-)definite matrix is denoted by $A\succ 0 \; (A \succeq 0)$. The Frobenius norm of a matrix $A \in \mathbb{R}^{m \times n}$ is denoted by $\|A\|_{\mathrm{F}} = \sqrt{\operatorname{trace}(A^\top A)}.$
\section{Problem Formulation}\label{sec:problem}
In this section, we describe the process, the data-driven controller, and the adversary.
\subsection{Problem setup}
Consider a process whose dynamics are represented by
\begin{equation}\label{P}
    x[k+1] = Ax[k] + Bu[k], \quad x[0]=0
\end{equation}
where $x[k] \in \mathbb{R}^n$ denotes the physical state of the process, $u[k] \in \mathbb{R}^m$ denotes the applied control input, and the matrices are of appropriate dimensions.
\begin{assumption}\label{ass:ctrb}
The pair $(A,B)$ is controllable. \hfill $\triangleleft$
\end{assumption}
We consider an operator who does not have access to the matrices $A$ and $B$, and who aims to design a feedback control policy for the process \eqref{P}. To this end, the operator employs data-driven control techniques \citep{de2019formulas} and applies persistently exciting (PE) inputs $u[k]$ to the process. The corresponding state measurements $x[k]$ are transmitted to the operator through a communication network that may be subject to cyber-attacks.

In this paper, we consider an adversary capable of injecting false data into the measurement channel, modeled as
\begin{equation}\label{eq:xa}
    \tilde{x}[k] = x[k] + a[k],
\end{equation}
where $a[k]$ is the attack signal injected by the adversary. The data collected by the operator is then denoted as
\begin{equation}\label{Dc}
    \mathcal{D}_c \triangleq \bigcup_{k=t_1}^{t_2} \{u[k], \tilde{x}[k]\}, 
    \qquad T \triangleq t_2 - t_1,
\end{equation}
where $t_1$ is the time index at which the operator begins applying PE inputs, and $T$ denotes the length of the dataset. Without loss of generality, we assume $t_1 = 0$. Since $\mathcal{D}_c$ is used by the operator to learn a stabilizing controller, inspired by common terminology in machine learning, we refer to $\mathcal{D}_c$ as the \emph{training dataset}.
\begin{assumption}
The training dataset $\mathcal{D}_c$ satisfies 
\begin{equation}\label{ineq:T}
    T \ge (m+1)n+m. \qquad \hfill \triangleleft
\end{equation}
\end{assumption}
Condition \eqref{ineq:T} ensures that the operator is able to apply PE inputs, which is necessary for designing stabilizing controllers \citep{de2019formulas}.

To detect potential attacks, the operator employs a detector of the form
\begin{equation}\label{eq:alarm}
\frac{\| W \tilde{x} \|_{2,[0,T]}}{\| u \|_{2,[0,T]}} > \gamma \;\; \implies \;\; \text{alarm},
\end{equation}
where $\gamma > 0$ is a detection threshold and $W$ is a weighting matrix. In essence, an alarm is raised when the finite-horizon $\ell_2$ gain of the plant exceeds a bound. The threshold $\gamma$ may be tuned by the operator using prior knowledge of the process around a different operating point. Moreover, if short segments of attack-free data or data collected from different attack-free operating points are available, they can be used to select an appropriate threshold \cite[Section~4]{NGUYEN2025169}. If the operator does not have access to any such data or prior information about the plant, $\gamma$ may be chosen to be a sufficiently large positive number; the results of this paper remain valid under this choice.

Many studies analyze stealthy attacks using model-based detectors (see Table~2 in \cite{anand2025quantifying}), but such detectors are unsuitable here since the operator does not know $A$ or $B$. Our detector, however, is structurally similar to that in \cite{teixeira2015strategic}.
\subsection{Controller description}
In this paper, we consider two different control objectives
\begin{enumerate}
    \item In Section~\ref{sec:design}, the operator employs a data-driven technique to design a stabilizing feedback gain \citep[Theorem 3]{de2019formulas}. 
    \item In Section~\ref{sec:design2}, the operator employs a data-driven technique to design an $H_2$-optimal feedback gain \citep[Theorem 4]{de2019formulas}. 
\end{enumerate}

In Section~\ref{sec:design}, we present an attack policy capable of inducing the operator to learn an {unstable} controller. In Section~\ref{sec:design2}, we present an attack policy capable of causing the operator to learn a {suboptimal} controller with a higher closed-loop $H_2$ cost. 
%
\subsection{Adversarial description}\label{sec:adv:res}
As described earlier, we consider an adversary that injects false data into the measurement channels. In line with \cite{teixeira2015secure}, we detail the resources and objectives of the adversary below

\emph{Disclosure resources:} The adversary can eavesdrop on both actuator and sensor channels. 
 
 \emph{Disruption resources:} The adversary can inject false data into the sensor channels. 
 
 \emph{Adversarial constraints:} In order to analyze the worst-case effect of attacks, we consider an adversary that injects an attack signal without raising any alarms. We define such attacks as stealthy attacks.
  	\begin{definition}
    	An attack signal is defined to be stealthy if it satisfies $\frac{ \| W \tilde{x} \|_{2,[0,T]} }{ \| {u} \|_{2,[0,T]} } \leq \gamma$. $\hfill \triangleleft$
    	\end{definition}
 \emph{Adversarial objectives:} We consider two different objectives for the adversary
    \begin{enumerate}
        \item In Section~\ref{sec:design}, the adversary injects false data to cause the operator to learn an {unstable} feedback gain. 
        \item In Section~\ref{sec:design2}, the adversary injects false data to cause the operator to learn a {suboptimal} feedback gain (with respect to the $H_2$-optimal controller).
    \end{enumerate}
\section{Attack policy against data-driven stabilization}\label{sec:design}
In this section, we first provide a recap of data-driven stabilizing control design algorithm. We then present the corresponding attack policy that renders the controller unstable, along with an attack-design algorithm. Finally, we provide a numerical example to illustrate our results.
\subsection{Preliminaries: data-driven stabilization}
From \cite{de2019formulas}, we next recall the result for designing a state-feedback controller from $\Dc_c$.
\begin{prop}\label{lem:SFC}
Let $a[k]=0, \forall k \in \mathbb{Z}^+$ and let $u_{[0,T]}$ in $\Dc_c$ be PE of order $n+1$. Then any controller of the form
\begin{equation}\label{eq:K}
K = U_{0,1,T}Q(\tilde{X}_{0,T}Q)^{-1}
\end{equation}
stabilizes the closed-loop system, i.e., $\max_i |\lambda_i(A-BK)| < 1$, where $Q \in \mathbb{R}^{T \times n}$ is any matrix that satisfies
\begin{equation}\label{eq:LMI}
    \begin{bmatrix}
        \tilde{X}_{0,T}Q & \tilde{X}_{1,T}Q\\
        Q^\top \tilde{X}_{1,T}^\top & \tilde{X}_{0,T}Q
    \end{bmatrix} \succ 0.
\end{equation}
Here, $\tilde{X}_{1,T}$ and $\tilde{X}_{0,T}$ are Hankel matrices generated from the measurements in $\Dc_c$. \hfill $\square$
\end{prop}
Given that the operator employs the design procedure in Lemma~\ref{lem:SFC}, we now address the following question

\say{How can the adversary design a stealthy attack policy $a[k]$ such that the operator learns an unstable controller $\tilde{K}$, i.e., $\max_i |\lambda_i(A - B\tilde{K})| > 1$?}
\subsection{Destabilizing attack policy}
Let us consider that the adversary designs a controller $\tilde{K}$ that makes the closed-loop system unstable. Next, the adversary aims to design an attack policy $a[k]$ such that the solution to \eqref{eq:K} becomes $\tilde{K}$. However, from \eqref{eq:K} and \eqref{eq:LMI}, it can be observed that the controller gain $K$ in \eqref{eq:K} is non-unique because the solution $Q$ in \eqref{eq:LMI} is not unique. For instance, if $Q_1$ is a solution to \eqref{eq:LMI}, then $\alpha Q_1$ with any $\alpha < -1$ is also a solution. Thus, the adversary cannot guarantee that the resulting controller gain in \eqref{eq:K} is exactly $\tilde{K}$. However, the adversary can ensure that $\tilde{K}$ is a feasible controller gain when solving \eqref{eq:K}–\eqref{eq:LMI}. We next formally define the notion of feasibility.

\begin{definition}[$\Dc_c$ feasibility]
A controller gain $\tilde{K}$ is said to be $\Dc_c$-\emph{feasible} if there exists a matrix $Q$ such that 
\begin{equation}\label{eq:K:feasibility}
\tilde{K} = U_{0,1,T}Q(\tilde{X}_{0,T}Q)^{-1},
\end{equation}
where $Q$ satisfies \eqref{eq:LMI}, and the Hankel data matrices are derived from $\Dc_c$. $\hfill \triangleleft$
\end{definition}

In other words, when the operator solves for a data-driven controller using \eqref{eq:K} - \eqref{eq:LMI}, if the resulting controller gain could possibly be $\tilde{K}$, then $\tilde{K}$ is $\Dc_c$-feasible. The objective of this section is therefore to show how the adversary can construct an attack policy such that $\tilde{K}$ becomes $\Dc_c$-feasible. Using the above definition, we now propose a method to generate the attacked measurements $\tilde{x}$.
\begin{theorem}\label{thm:1}
Let the operator apply PE inputs $u[k]$, and let $\tilde{K} \in \mathbb{R}^{m \times n}$ be the adversary's unstable controller gain with elements $\tilde{k}_{ij}$. Define $ V = \begin{bmatrix}v_1 & \dots & v_n \end{bmatrix}, \quad v_j = \frac{1}{m} \sum_{i=1}^m \tilde{k}_{ij}.$ Let the attack signal in \eqref{eq:xa} be generated by
%
%
\begin{align}\label{eq:fake:system}
x_a[k+1] &= \tilde{A} x_a[k] + \tilde{B} u[k], \quad x_a[0] =0,\\
a[k] &= -x[k] + x_a[k],
\end{align}
\begin{equation}\label{mat:tilde}
\left[\begin{array}{c|c} \tilde{A} & \tilde{B} \end{array}\right] =
\left[
\begin{array}{ccc|c}
0_{(n-1) \times 1} & & \kappa I_{n-1} & {\color{white}\frac{\kappa}{m}}0_{(n-1) \times m} \\
\noalign{\vskip 0.2cm} \hline \noalign{\vskip 0.2cm}
& \kappa V & & \frac{\kappa}{m} 1_{1 \times m}
\end{array}
\right].
\end{equation}
Then
\begin{enumerate}[label=(\alph*)]
    \item $\tilde{K}$ is $\Dc_c$-feasible for the operator.
    \item There exists $\kappa \in (0,1]$ such that the attack \eqref{eq:fake:system} is stealthy.
\end{enumerate}
\end{theorem}
\begin{pf}
Let $\bar{A},\bar{B}$ be the matrices $(\tilde{A},\tilde{B})$ when $\kappa=1$, i.e., $(\tilde{A},\tilde{B}) = (\kappa\bar{A},\kappa\bar{B})$. Let $\hat{B}$ be the matrix such that $\bar{B}= \frac{1}{m}\hat{B}$. Now we are ready to present the proofs.\\
\emph{Proof of $(a)$:}  Observe that when $m-1$ inputs are removed from $\hat{B}$ (that is, only one control input is available), we obtain a single column $b = [0, 0, \ldots, 0, 1]^T$. The pair $(\bar{A}, b)$ forms the standard SISO controllable canonical form, which is always controllable. 

Since adding inputs cannot reduce controllability (especially in this case where the inputs are added along the same direction), the system $(\bar{A}, \hat{B})$ is controllable.

If the tuple $(\bar{A}, \hat{B})$ is controllable, it also follows that $(\bar{A}, \frac{1}{m}\hat{B})$ is controllable.To see this, note that the controllability matrix of $(\bar{A}, \frac{1}{m}\hat{B})$ is $W_{\bar{B}} =$
\begin{equation}
\begin{bmatrix} \frac{1}{m}\hat{B} & \;\bar{A}\left(\frac{1}{m}\hat{B}\right) & \;\bar{A}^2\left(\frac{1}{m}\hat{B}\right) &\; \cdots &\; \bar{A}^{n-1}\left(\frac{1}{m}\hat{B}\right) \end{bmatrix} = \frac{1}{m}W_{\hat{B}}
\end{equation}
where $W_{\hat{B}}$ is the controllability matrix of the tuple $(A,\hat{B})$. Since $\frac{1}{m} \neq 0$ and $\mathrm{rank}(W_{\hat{B}}) = n$, we have $\mathrm{rank}(W_{\bar{B}}) = n$. Thus $(\bar{A}, \bar{B})$ is controllable.

The controllability matrix of $(\tilde{A}, \tilde{B})$ is
\begin{equation}
W_{\tilde{B}} = W_{\bar{B}} \cdot \mathrm{diag}([\kappa, \kappa^2, \kappa^3, \ldots, \kappa^n]).
\end{equation}
Since $\kappa \neq 0$, the diagonal matrix $\mathrm{diag}([\kappa, \kappa^2, \kappa^3, \ldots, \kappa^n])$ is invertible. Therefore $\mathrm{rank}(W_{\tilde{B}}) = n$. Then it follows that $W_{\tilde{B}}$ is full rank, and the tuple $(\tilde{A}, \tilde{B})$ is controllable.

Since $(\tilde{A},\tilde{B})$ is controllable, from Theorem~1 in \cite{de2019formulas}, we know that if a controller ${K}$ stabilizes the tuple $(\tilde{A},\tilde{B})$, it can be equivalently written of the form \eqref{eq:K:feasibility} where $Q$ is obtained from \eqref{eq:LMI} and Hankel data matrices are derived from $\Dc_c$. Thus, if we show that $\tilde{K}$ stabilizes the tuple $(\tilde{A},\tilde{B})$, the proof concludes. To this end, we derive the following $\tilde{A}-\tilde{B}\tilde{K} =$
\begin{equation}
 \begin{bmatrix}
        0_{n-1 \times 1} &\; &  \kappa I_{n-1} \\
 \;& \kappa V & \;
    \end{bmatrix} - \begin{bmatrix}
        0_{n-1 \times n}\\
        \kappa V
    \end{bmatrix} = \begin{bmatrix}
        0_{n-1 \times 1} & \;&  \kappa I_{n-1} \\
 \;& 0_{1 \times n} & \;
    \end{bmatrix}
\end{equation}
Since the matrix $ \tilde{A}-\tilde{B}\tilde{K}$ is upper triangular with zero entries on the diagonal, it holds that $\max_i|{\lambda}_i( \tilde{A}-\tilde{B}\tilde{K})| <1$. Thus, the matrix $ \tilde{A}-\tilde{B}\tilde{K}$ is Schur stable, which concludes the proof. 

\emph{Proof of $(b)$:} The attack policy \eqref{eq:fake:system} along with \eqref{P} and \eqref{eq:xa} can be written as 
\begin{align}
x_a[k+1] &= \tilde{A} x_a[k] + \tilde{B} u[k], x_a[k] =0\\
\tilde{x}[k] &= x_a[k]
\end{align}
Then, proving $(b)$ is equivalent to showing that any given $\gamma > 0$, there exists $\kappa > 0$ such that
\begin{equation}
\gamma(\tilde{A}, \tilde{B}, W; T) = \sup_{u \neq 0} \frac{\|\tilde{y}\|_{2,[0,T]}}{\|u\|_{2,[0,T]}} < \gamma,
\end{equation}
where $\tilde{y}[k] = W\tilde{x}[k]$ and $W$ is the weight matrix in \eqref{eq:alarm}. For the system $(\tilde{A}, \tilde{B}, W)$, the output with input $u$ is
\begin{equation}
\tilde{y}[k] = W\sum_{i=0}^{k-1} (\kappa \bar{A})^{k-1-i} (\kappa \bar{B}) u[i] = \kappa W \sum_{i=0}^{k-1} (\kappa \bar{A})^{k-1-i} \bar{B} u[i].
\end{equation}
Taking the norm
\begin{equation}
\|\tilde{y}[k]\|^2 = \kappa^2 \left\| W \sum_{i=0}^{k-1} \kappa^{k-1-i} \bar{A}^{k-1-i} \bar{B} u[i]\right\|^2.
\end{equation}
Since $0 < \kappa < 1$, we have $\kappa^{k-1-i} \leq 1$ for all $i \in \{0, \ldots, k-1\}$. Therefore
\begin{equation}
\|\tilde{y}[k]\|^2 \leq \kappa^2 \left\| W \sum_{i=0}^{k-1} \bar{A}^{k-1-i} \bar{B} u[i]\right\|^2 = \kappa^2 \|\bar{y}[k]\|^2,
\end{equation}
where $\bar{y}[k] = W \displaystyle \sum_{i=0}^{k-1} \bar{A}^{k-1-i} \bar{B} u[i]$ is the output of the unscaled system $(\bar{A}, \bar{B})$ with the same input $u$. Summing over the time horizon $[0, T]$
\begin{equation}
\|\tilde{y}\|_{2,[0,T]}^2 = \sum_{k=0}^{T-1} \|\tilde{y}[k]\|^2 \leq \kappa^2 \sum_{k=0}^{T-1} \|\bar{y}[k]\|^2 = \kappa^2 \|\bar{y}\|_{2,[0,T]}^2.
\end{equation}
Taking square roots, it holds that
\begin{equation}
\frac{\|\tilde{y}\|_{2,[0,T]}}{\|u\|_{2,[0,T]}} \leq \kappa \frac{\|\bar{y}\|_{2,[0,T]}}{\|u\|_{2,[0,T]}}.
\end{equation}
Taking the supremum over all inputs
\begin{equation}
\gamma(\kappa\bar{A}, \kappa\bar{B}, W; T) = \sup_{u \neq 0} \frac{\|\tilde{y}\|_{2,[0,T]}}{\|u\|_{2,[0,T]}} \leq \kappa \cdot \gamma(\bar{A}, \bar{B}, W; T).
\end{equation}
Assuming $\gamma(\bar{A}, \bar{B}, W; T) > 0$, choose $ \kappa = \frac{\gamma}{2\gamma(\bar{A}, \bar{B}, W; T)}$. Then
\begin{equation}
\gamma(\kappa\bar{A}, \kappa\bar{B}, W; T) \leq \kappa \cdot \gamma(\bar{A}, \bar{B}, W; T) = \frac{\gamma}{2} < \gamma.
\end{equation}
This completes the proof. $\hfill \blacksquare$
\end{pf}

We have now shown that if the adversary generates the attacked measurements using \eqref{eq:xa} and \eqref{eq:fake:system}, then $\tilde{K}$ is a feasible controller gain for the operator. As a result, the operator may learn the unstable controller $\tilde{K}$. Using the results in Theorem~\ref{thm:1}, we also provide an algorithm to generate the stealthy attack policy in Algorithm~\ref{alg:1}. 

Once the unstable controller is deployed, the adversary can continue sending attacked measurements generated by \eqref{eq:fake:system} to avoid detection. However, in reality, the process will behave poorly. We illustrate these results next through a numerical example.

\begin{algorithm}[t]
\caption{Stealthy De-stabilizing Attack Policy}
\label{alg:1}
\begin{algorithmic}[1]
\STATE \textbf{Input:} $n$, $\gamma$, $T$, and convergence threshold $\epsilon$.
\STATE \textbf{Initialization:} Pick an unstable controller gain $\tilde{K}$.
\STATE Construct matrices $\tilde{A}$ and $\tilde{B}$ as functions of $\kappa$ in \eqref{mat:tilde}.
\STATE Perform line search over $\kappa$ (see also Remark~\ref{rem:kappa}):
\begin{equation}
\kappa^* = \arg\min_{\kappa} \left| \left(\sup_{u \neq 0}\frac{\|W\tilde{x}\|_{\ell_2,[0,T]}}{\|u\|_{\ell_2,[0,T]}}\right) - \gamma \right|.
\end{equation}
\STATE \textbf{Return:} The matrices $\tilde{A}$ and $\tilde{B}$ constructed using $\kappa^*$.
\end{algorithmic}
\end{algorithm}

\begin{rem}\label{rem:kappa}
From the proof of part (b) in Theorem~\ref{thm:1}, a closed-form choice of $\kappa$ that guarantees $\frac{\|W\tilde{x}\|_{\ell_2,[0,T]}}{\|u\|_{\ell_2,[0,T]}} \le \gamma$ is $\kappa = \frac{\gamma}{2\delta}$, where $\delta = \sup_{u \neq 0}\left[ \frac{\|W\tilde{x}\|_{2,[0,T]}}{\|u\|_{2,[0,T]}} \Bigg|_{\kappa=1} \right].$ In other words, $\delta$ is the finite-horizon $\ell_2$ gain of the system $(\tilde{A},\tilde{B},W)$ with $\kappa=1$. $\hfill \triangleleft$
\end{rem}
\subsection{Numerical example}
\begin{figure}[t]
    \centering
    \includegraphics[width=8cm]{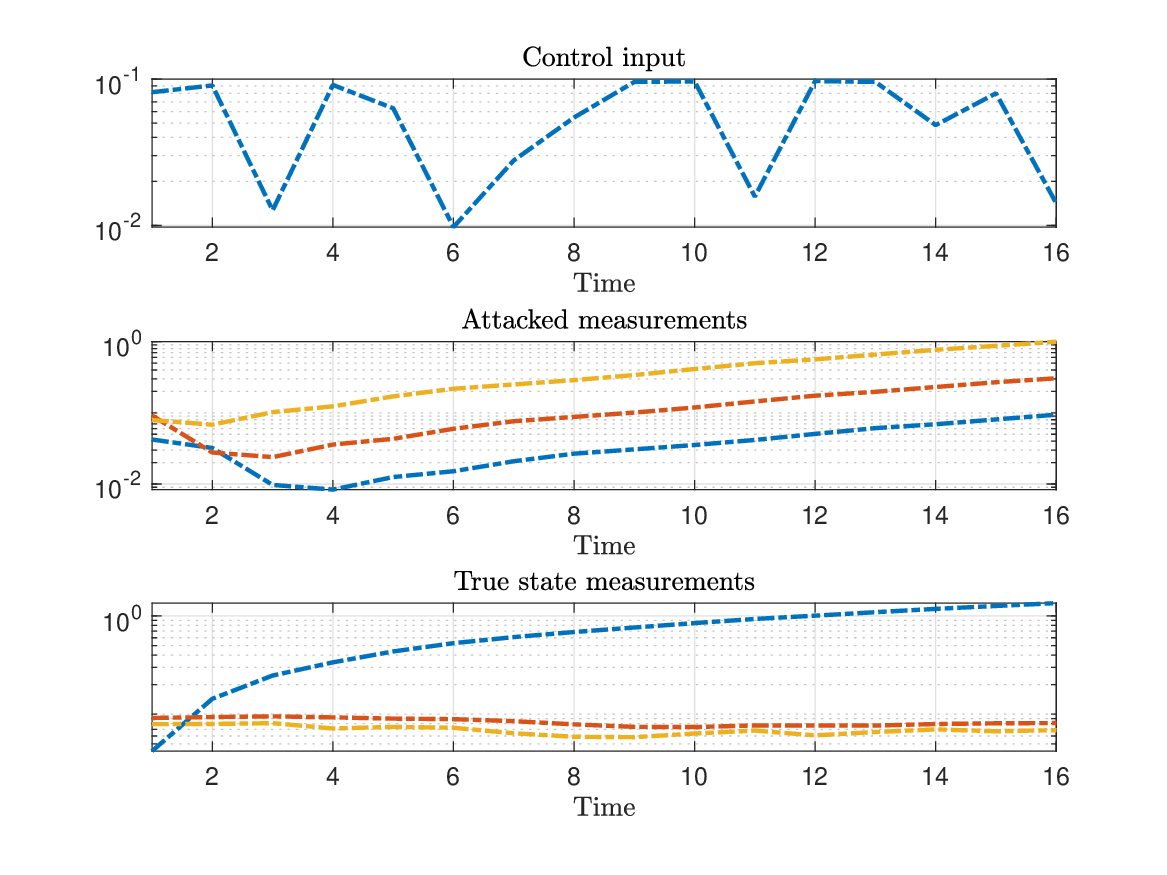}
    \caption{(Top) PE inputs applied of length $T$. (Middle) Attacked measurements generated by \eqref{eq:NE2}. (Bottom) True measurements generated by the process \eqref{eq:NE1}.}
    \label{fig:NE1}
\end{figure}

Consider the continuous-time dynamical system
\begin{equation}\label{eq:NE1}
\dot{x}(t) = 
\begin{bmatrix}
-0.1 & 3 & 4\\
0 & -5 & 6\\
0 & 0 & -1
\end{bmatrix}x(t) +
\begin{bmatrix}
1\\0\\1
\end{bmatrix}u(t).
\end{equation}
We discretize \eqref{eq:NE1} using a zero-order hold with a sampling time of $T_s = 0.15\;\mathrm{s}$ to obtain the dynamics in \eqref{P}. The adversary then designs an unstable controller
\begin{equation}\label{eq:K:unstable}
    \tilde{K} =
    \begin{bmatrix}
    -0.01 & -2.67 & 3.27
    \end{bmatrix}.
\end{equation}
Using Theorem~\ref{thm:1}, the adversary generates attacked measurements according to \eqref{eq:fake:system}, where
\begin{equation}\label{eq:NE2}
\tilde{A} = 
\begin{bmatrix}
0 & \kappa & 0\\
0 & 0 & \kappa\\
-0.01\kappa & -2.67\kappa & 3.27\kappa
\end{bmatrix},\;
\tilde{B} = 
\begin{bmatrix}
0\\0\\\kappa
\end{bmatrix},\; \kappa=1.
\end{equation}
To design a data-driven controller, the operator applies PE inputs of length $T=16$. The applied input, the plant response, and the attacked measurements generated by the adversary are shown in Fig.~\ref{fig:NE1}. We can see from Fig.~\ref{fig:NE1} that the true state measurement, and the attacked measurement lie in the same range.

\begin{figure}[t]
    \centering
    \includegraphics[width=8cm]{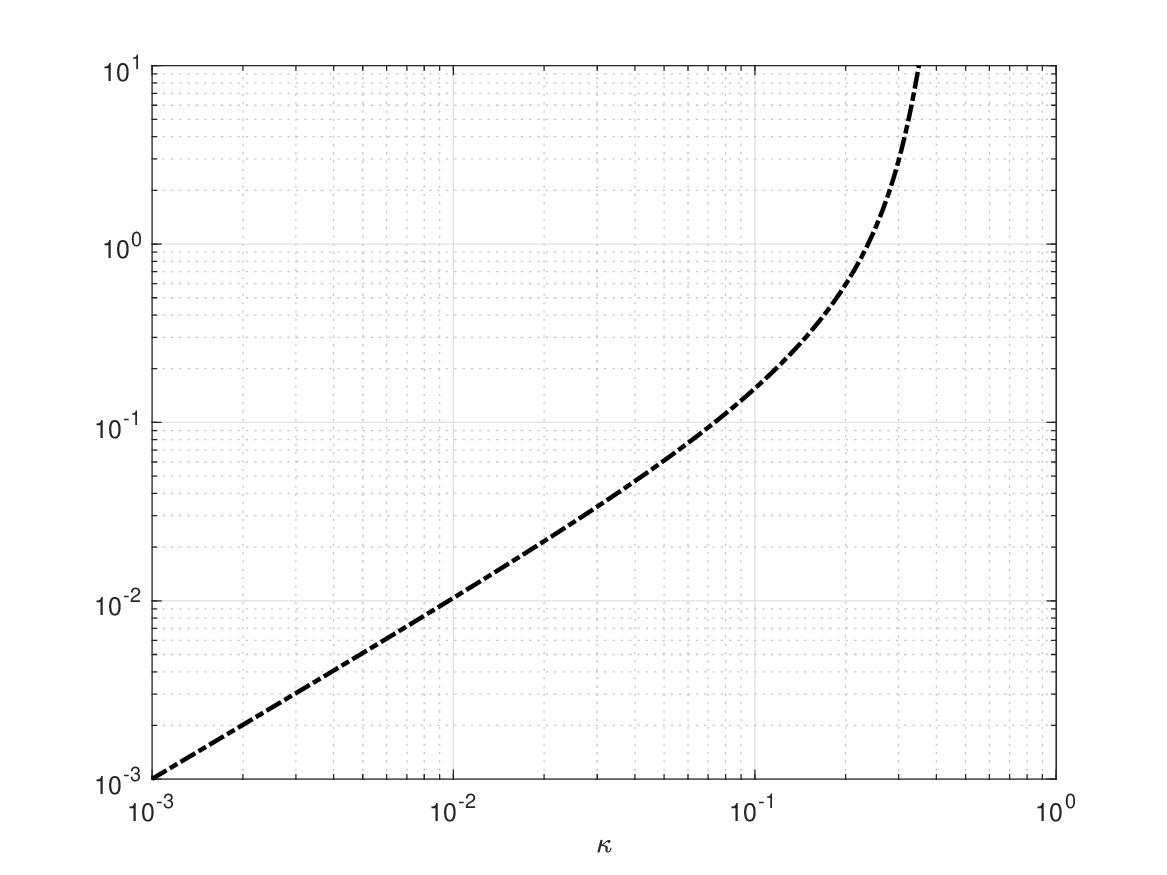}
    \caption{Value of $\displaystyle \sup_{u \neq 0}\frac{\|W\tilde{x}\|_{\ell_2,[0,T]}}{\|u\|_{\ell_2,[0,T]}}$ for varying values of $\kappa$.}
    \label{fig:NE1a}
\end{figure}

The value of $\sup_{u \neq 0}\frac{\|W\tilde{x}\|_{\ell_2,[0,T]}}{\|u\|_{\ell_2,[0,T]}}$ for varying $\kappa$ is depicted in Fig.~\ref{fig:NE1a}, where we set $W = I_3$. From Fig.~\ref{fig:NE1a}, we observe that the attack can be scaled for stealthiness for any value of $\gamma$. For very small values of $\gamma$, the outputs are scaled accordingly. However, for cases where the CT process \eqref{eq:NE1} has a pole close to zero ($p=-0.1$), the applied inputs will be small in magnitude ($u\approx 10^{-1}$). Consequently, the scaled value of $\kappa$ also yields small output values, which should not surprise the operator. Next, we consider an adversary targeting the $H_2$-optimal controller.
\section{Attack policy against data-driven $H_2$-optimal control}\label{sec:design2}
In this section, we first recap the data-driven $H_2$-optimal control design algorithm. We then present the corresponding attack policy that renders the data-driven controller suboptimal. Finally, we provide a numerical example to illustrate the results.
\subsection{Preliminaries: data-driven linear quadratic regulation}
Consider the process \eqref{P} with the virtual performance signal denoted as follows
\begin{equation}\label{eq:LQR}
\begin{aligned}
x[k+1] &= Ax[k] + Bu[k] + \eta[k],\\
z[k] &= 
\begin{bmatrix}
Q_x^{1/2} & 0\\[1mm]
0 & R^{1/2}
\end{bmatrix}
\begin{bmatrix}
x[k]\\
u[k]
\end{bmatrix},
\end{aligned}
\end{equation}
where $\eta[k]$ is white noise, $z[k]$ is the performance output, and $Q_x \succeq 0$, $R \succ 0$. 

In the $H_2$-optimal control problem, the objective is to design a state-feedback controller that minimizes the effect of $\eta$ on $z$. Equivalently, the operator seeks to solve
\[
J^* \triangleq \min_{K} \|h\|_2, \qquad h: \eta \mapsto z.
\]
From \cite{de2019formulas}, we recall the data-driven construction of the $H_2$-optimal controller from $\Dc_c$.
\begin{prop}\label{lem:H2}
Let $a[k]=0, \forall \; k \in \mathbb{Z}^+$, and let $u_{[0,T]}$ in $\Dc_c$ be PE of order $n+1$. Then the $H_2$-optimal controller for \eqref{eq:LQR} can be computed as
\[
\tilde{K} = U_{0,1,T} Q\, (\tilde{X}_{0,T} Q)^{-1},
\]
where $Q$ solves
\begin{equation}\label{eq:H2}
\begin{aligned}
\min_{Q, X} \quad & \mathrm{trace}(Q_x \tilde{X}_{0,T} Q) + \mathrm{trace}(X) \\
\text{subject to} \quad &
\begin{bmatrix}
X & R^{1/2} U_{0,1,T} Q \\
Q^\top U_{0,1,T}^\top R^{1/2} & \tilde{X}_{0,T} Q
\end{bmatrix} \succeq 0, \\
&
\begin{bmatrix}
\tilde{X}_{0,T} Q - I_n & \tilde{X}_{1,T} Q \\
Q^\top \tilde{X}_{1,T}^\top & \tilde{X}_{0,T} Q
\end{bmatrix} \succeq 0.
\end{aligned}
\end{equation}
The Hankel matrices are constructed from $\Dc_c$. \hfill $\square$
\end{prop}
When $a[k]=0$, denote by $J^*$ the $H_2$ cost achieved by the optimal controller. As we show later, an adversary can design an attack policy that increases this cost during the controller-design phase. Let $J_a$ denote the cost under attack. We now address the following question

\say{Given $u[k]$, how can the adversary design an attack policy $a[k]$ such that $J_a \gg J^*$?}
\subsection{Attack policy}
Before presenting the results, we introduce some notation. As depicted in \eqref{eq:xa}, let $a[k],\, k \in \{0,1,\dots,T\}$ denote the attack injected by the adversary. The corresponding Hankel matrices are denoted by $A_{0,T}$ and $A_{1,T}$ (analogous to $X_{0,T}$ and $X_{1,T}$). Then,
\begin{equation}
\tilde{X}_{0,T} = X_{0,T} + A_{0,T},\qquad 
\tilde{X}_{1,T} = X_{1,T} + A_{1,T}.
\end{equation}
The adversary seeks to inject an attack sequence that increases the operator's $H_2$-optimal cost. Since the operator solves \eqref{eq:H2} to obtain the controller, we design an adversary that chooses the attack by solving
\begin{equation}\label{H2:attack}
\begin{aligned}
\max_{\mathcal{A}} \;\;\min_{Q, X} \quad & \mathrm{trace}(Q_x \tilde{X}_{0,T} Q) + \mathrm{trace}(X) \\
\text{subject to} \quad &
\begin{bmatrix}
X & R^{1/2} U_{0,1,T} Q \\
Q^\top U_{0,1,T}^\top R^{1/2} & \tilde{X}_{0,T} Q
\end{bmatrix} \succeq 0, \\
&
\begin{bmatrix}
\tilde{X}_{0,T} Q - I_n & \tilde{X}_{1,T} Q \\
Q^\top \tilde{X}_{1,T}^\top & \tilde{X}_{0,T} Q
\end{bmatrix} \succeq 0, \\
&\;\Vert W \tilde{X}_{0,T}\Vert_F^2 \leq \gamma \Vert U_{0,T}\Vert_F^2,
\end{aligned}
\end{equation}
where $\mathcal{A}=\{a[0],a[1],\dots,a[T]\}$. Here the stealthiness constraint is reformulated in terms of Frobenius norms by considering the attack trajectory. In other words,
\begin{equation}
\|W\tilde{x}\|_{2,[0,T]}^2 = \sum_{k=0}^T \|W\tilde{x}[k]\|_2^2 = \|W\tilde{X}_{0,T}\|_F^2
\end{equation}
Similarly, $\|u\|_{2,[0,T]}^2 = \|U_{0,T}\|_F^2$. Therefore, the stealthiness constraint $\frac{ \| W \tilde{x} \|_{2,[0,T]} }{ \| {u} \|_{2,[0,T]} } \leq \gamma$ is equivalent to $\|W\tilde{X}_{0,T}\|_F \leq \gamma \|U_{0,T}\|_F$.

The problem \eqref{H2:attack} is non-convex since the objective function and the constraints are bilinear in the decision variables. Thus, solving it exactly is difficult, and we adopt a suboptimal alternating-minimization procedure, summarized in Algorithm~\ref{alg:1}. We next show that under such attacks, the closed-loop $H_2$ cost strictly increases.
\begin{theorem}\label{lem:H2:1}
Consider system \eqref{eq:LQR} with $x[0]=0$, and let $u_{[0,T]}$ be PE of order $n+1$. Let $J^*$ denote the optimal ${H}_2$ cost of the closed-loop system when the controller is obtained by solving \eqref{lem:H2} using the attack-free data. Suppose the measurements are corrupted by a non-zero attack, i.e., $\exists k \in \{0,1,\dots,T\}$ such that $a[k] \neq 0$. Let $J_a$ denote the ${H}_2$ cost of the closed-loop system when the controller is obtained from the attacked data. Then $J_a > J^*$.
\end{theorem}
\begin{pf}
Since $(A,B)$ is controllable (Assumption~\ref{ass:ctrb}) and the cost matrices satisfy $Q_x \succeq 0$ and $R \succ 0$, the ${H}_2$-optimal control problem has a strictly convex cost function $J(K)$ with a unique minimizer $K^*$ \cite[Chapter 10.3.4]{monnier2024vda}. 

For the true system \eqref{P} with zero initial conditions, a given input sequence $u_{[0,T]}$ uniquely determines the corresponding state trajectories $x_{[1,T+1]}$. When the measurements are corrupted by a non-zero attack, the perturbed data no longer satisfies the true system dynamics. Solving the data-driven ${H}_2$ problem \eqref{eq:H2} with corrupted data yields a controller $K_a$ associated with the perturbed trajectories, and thus $K_a \neq K^*$. By strict convexity of $J(K)$ and uniqueness of $K^*$, we obtain
\[
J_a = J(K_a) > J(K^*) = J^*,
\]
which completes the proof. \hfill $\blacksquare$
\end{pf}
Thus we have shown that for any non-zero attack, the closed-loop $H_2$ cost increases and we also provided an algorithm to increase the cost. In the next section, we provide additional discussion related to Algorithm~\ref{alg:1}.

\begin{algorithm}[t]
\caption{Stealthy attack design algorithm against data-driven $H_2$-optimal control}
\label{alg:1}
\label{alg:alternating_min}
\begin{algorithmic}[1]
\STATE \textbf{Input:} Attack-free trajectory data $(X_{0,T}, X_{1,T}, U_{0,T})$, weights $Q_x, R$, and maximum iterations $N_{\max}$
\STATE \textbf{Output:} Attack matrices $A_{0,T}, A_{1,T}$
\STATE \textbf{Initialize:} $\tilde{X}_{0,T}^{(0)} \gets X_{0,T}$, $\tilde{X}_{1,T}^{(0)} \gets X_{1,T}$
\FOR{$k = 1, 2, \ldots, N_{\max}$}
    \STATE \textbf{Step 1: Minimize over $(Q, X)$} with fixed $\tilde{X}_{0,T}^{(k-1)}, \tilde{X}_{1,T}^{(k-1)}$
    \STATE Solve:
    \begin{align*}
    \min_{Q, X} \quad & \mathrm{trace}(Q_x \tilde{X}_{0,T}^{(k-1)} Q) + \mathrm{trace}(X) \\
    \text{s.t.} \quad &
    \begin{bmatrix}
    X & R^{1/2} U_{0,T} Q \\
    Q^\top U_{0,T}^\top R^{1/2} & \tilde{X}_{0,T}^{(k-1)} Q
    \end{bmatrix} \succeq 0 \\
    &
    \begin{bmatrix}
    \tilde{X}_{0,T}^{(k-1)} Q - I_n & \tilde{X}_{1,T}^{(k-1)} Q \\
    Q^\top (\tilde{X}_{1,T}^{(k-1)})^\top & \tilde{X}_{0,T}^{(k-1)} Q
    \end{bmatrix} \succeq 0
    \end{align*}
    \STATE Store solution: $Q^{(k)} \gets Q^*$, $X^{(k)} \gets X^*$
    \STATE Compute $K^{(k)} = (X_{0,T}Q^{(k)})^{-1}U_{0,T}Q^{(k)}$
    \STATE \textbf{Step 2: Maximize over $\mathcal{A}$} with fixed $Q^{(k)}, X^{(k)}$
    \STATE Solve:
    \begin{align*}
    \max_{\mathcal{A}} \;\; & \mathrm{trace}(Q_x \tilde{X}_{0,T} Q^{(k)}) + \mathrm{trace}(X^{(k)})  \\
    \text{s.t.} \;\; &
    \begin{bmatrix}
    X^{(k)} & R^{1/2} U_{0,T} Q^{(k)} \\
    (Q^{(k)})^\top U_{0,T}^\top R^{1/2} & \tilde{X}_{0,T} Q^{(k)}
    \end{bmatrix} \succeq 0 \\
    &
    \begin{bmatrix}
    \tilde{X}_{0,T} Q^{(k)} - I_n & \tilde{X}_{1,T} Q^{(k)} \\
    (Q^{(k)})^\top \tilde{X}_{1,T}^\top & \tilde{X}_{0,T} Q^{(k)}
    \end{bmatrix} \succeq 0 \\
    & \|W\tilde{X}_{0,T}\|_F \leq \gamma \|U_{0,T}\|_F \\
    & \tilde{X}_{0,T} = X_{0,T} + A_{0,T}, \, \tilde{X}_{1,T} = X_{1,T} + A_{1,T}
    \end{align*}
    \STATE Update: $\tilde{X}_{0,T}^{(k)} \gets X_{0,T} + A_{0,T}$, $\tilde{X}_{1,T}^{(k)} \gets X_{1,T} + A_{1,T}$
    \textbf{If}\;{$\Vert K^{(k)} - K^{(k-1)} \Vert < \epsilon$}, then \textbf{break}
\ENDFOR
\STATE \textbf{return} $A_{0,T} \gets \tilde{X}_{0,T}^{(K)} - X_{0,T}$, $A_{1,T} \gets \tilde{X}_{1,T}^{(K)} - X_{1,T}$
\end{algorithmic}
\end{algorithm}
\subsection{Discussion}\label{sec:discuss}
In the previous section, we presented an algorithm to design stealthy attacks against data-driven controllers to induce suboptimality. However, as discussed in \cite{de2019formulas}, the following rank condition
\begin{equation}\label{con:rank}
\text{rank} (\tilde{\Lambda})=n+m, \qquad 
\tilde{\Lambda} \triangleq 
\begin{bmatrix}
U_{0,1,T}\\ 
\tilde{X}_{0}
\end{bmatrix}
\end{equation}
is essential for the operator to design a controller. In other words, a data-driven $H_2$-optimal controller can be designed if and only if \eqref{con:rank} is satisfied. When there are no attacks, \eqref{con:rank} holds if the inputs are PE. Under attacks, however, the condition may no longer hold.

Thus, if the operator knows that the plant is linear, knows its order $n$, and applies PE inputs, then \eqref{con:rank} must be satisfied. If it is not, the operator may raise an alarm. In this section, we discuss two methods that enable the adversary to enforce this additional stealthiness.

\emph{Method 1:} One way to satisfy the rank condition \eqref{con:rank} is to include the rank constraint directly when solving the optimization problem in Step (10) of Algorithm~\ref{alg:1}. Because rank constraints are non-convex, they cannot be solved efficiently. However, convex surrogates for the rank function may be used, as suggested in \cite{fazel2001rank,recht2010guaranteed}.

\emph{Method 2:} Another approach is to restrict the class of admissible attacks so that the rank constraint is automatically met. In this method, we restrict the attack to be a constant bias injection \citep{tosun2024quickest} and derive necessary conditions under which \eqref{con:rank} holds.
\begin{theorem}
Let $u_{[0,T]}$ be PE of order $n+1$. Let the attack signal in \eqref{eq:xa} be a constant bias of the form $a[k] = \rho {1}_{n \times 1}$ with $\rho \in \mathbb{R}$. Then \eqref{con:rank} holds if 
\begin{equation}\label{eq:1}
{1}_{1 \times T} \notin \mathrm{RowSpace}( \Lambda), \qquad \Lambda \triangleq \begin{bmatrix}
U_{0,1,T}\\ 
X_{0}
\end{bmatrix}
\end{equation}
where $X_0$ is the Hankel matrix of the uncorrupted state measurements.
\end{theorem}
\begin{pf}
We proceed by contradiction. Suppose $\mathrm{rank}(\tilde{\Lambda}) < n+m$. Then the rows are linearly dependent, so there exist scalars $\alpha_1,\dots, \alpha_m \in \mathbb{R}$ and $\beta_1,\dots,\beta_n \in \mathbb{R}$, not all zero, such that
\[
\sum_{i=1}^m \alpha_i u_{i} + \sum_{j=1}^n \beta_j (x_j + \rho\, \mathbf{1}) = 0,
\]
where $u_i$ are the rows of $U_{0,1,T}$ and $x_j$ are the rows of $X_0$. Expanding,
\[
\sum_{i=1}^m \alpha_i u_{i} + \sum_{j=1}^n \beta_j x_j + \rho \sum_{j=1}^n \beta_j\, \mathbf{1} = 0.
\]
Define $v \triangleq \sum_{i=1}^m \alpha_i u_{i} + \sum_{j=1}^n \beta_j x_j$ and $\gamma \triangleq \sum_{j=1}^n \beta_j$. Then
\[
v + \rho\, \gamma\, \mathbf{1} = 0 \quad \Rightarrow \quad
v = -\rho\, \gamma\, \mathbf{1}.
\]
Since $v$ is a linear combination of the rows of $\Lambda$, we have $v \in \mathrm{RowSpace} (\Lambda)$.

\noindent\textbf{Case 1:} $\gamma \neq 0$.  
Then from $v = -\rho\, \gamma\, \mathbf{1}$, we obtain $ \mathbf{1} = -\frac{1}{\rho\, \gamma} v \in \mathrm{RowSpace} (\Lambda),$ contradicting assumption \eqref{eq:1}.

\noindent\textbf{Case 2:} $\gamma = 0$.  
Then $v = 0$, which implies $ \sum_{i=1}^m \alpha_i u_{i} + \sum_{j=1}^n \beta_j x_j = 0,$ with not all coefficients zero. Since $u_{[0,T]}$ is PE of order $n+1$, we have $
\mathrm{rank} (\Lambda)= n+m,$ so the rows are linearly independent. This is a contradiction. In both cases, we arrive at a contradiction. Therefore, \eqref{con:rank} must hold which concludes the proof. \hfill $\blacksquare$
\end{pf}
Thus, for constant-bias injection attacks, the adversary can check the simple condition \eqref{eq:1} to ensure that the rank condition is satisfied, allowing the operator to compute a feasible controller and therefore avoid raising an alarm. The same condition can also be used as a mitigation strategy to prevent such attacks. In other words, the operator should aim to design inputs such that \eqref{eq:1} does not hold. Then the adversary cannot inject attakcks whilst also ensuring feasibility. In the next section, we illustrate the results with a numerical example.
\subsection{Numerical example}\label{exmp:2}
Consider the continuous-time dynamical system
\begin{equation}\label{eq:NE2}
\dot{x}(t) = 
\begin{bmatrix}
-1 & 3 & 4\\
0 & -2 & 6\\
0 & 0 & -0.8
\end{bmatrix}x(t) +
\begin{bmatrix}
0.1\\0\\0.1
\end{bmatrix}u(t).
\end{equation}
We discretize \eqref{eq:NE2} using a bilinear transformation with sampling time $T_s = 0.01 \;\mbox{s}$ to obtain the system dynamics used in \eqref{P}. We select $Q_x = I_3$, $R = 1$, and $\gamma = 10^{\frac{3}{2}}$.

The optimal $H_2$ cost is $J^* = 61.7407$. The corresponding optimal state-feedback controller is 
\[
K^* = \begin{bmatrix}-0.5359 & -0.7937 & -3.0788\end{bmatrix}.
\]
We apply alternating minimization in Algorithm~\ref{alg:1} with $N_{\max} = 3$ iterations. Under the resulting attack, the closed-loop $H_2$ cost becomes $244.4900$, representing an approximate $400\%$ increase.

\begin{figure}
    \centering
    \includegraphics[width=8.4cm]{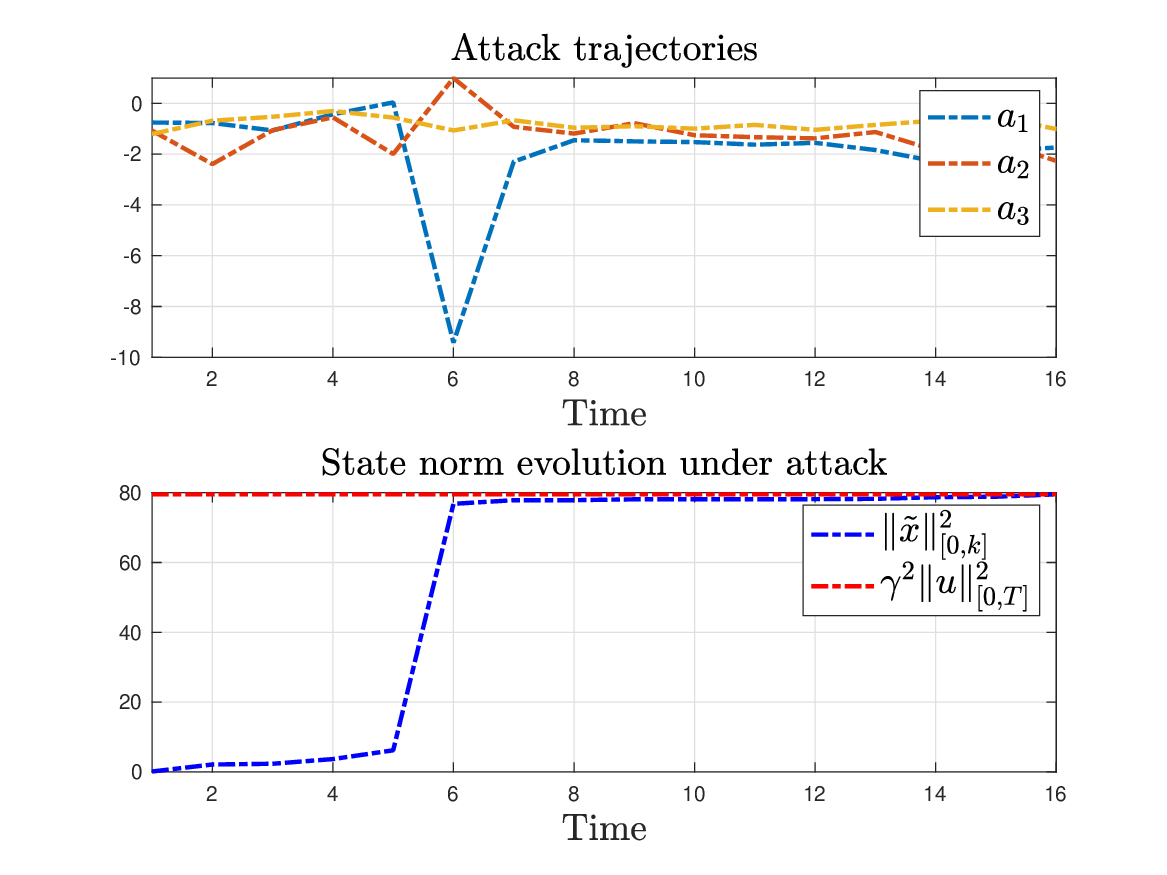}
    \caption{State trajectories, optimal attack signal, and detection threshold for Example~\ref{exmp:2}.}
    \label{fig:H2}
\end{figure}

Fig.~\ref{fig:H2} shows the optimal attack vector, the attacked state trajectories, and the detection threshold $\gamma \|u\|_{[0,T]}^2$. We observe that the attack is stealthy, remaining below the detection threshold throughout the horizon. Furthermore, the rank condition is satisfied under the computed attack, consistent with the discussion in Section~\ref{sec:discuss}.
\section{Conclusions and future works}\label{sec:con}
In this paper, we investigated the vulnerability of data-driven control systems to stealthy sensor attacks during the learning phase. We demonstrated that an adversary can manipulate sensor data to mislead the operator into learning an {unstable} state-feedback controller, thereby compromising system stability. For data-driven ${H}_2$ control, we formulated attacks that degrade closed-loop performance while ensuring that the operator can always compute a feasible controller. Our results highlight the critical importance of considering adversarial manipulations when designing data-driven controllers, particularly in scenarios where the operator has no prior knowledge of the plant dynamics. Future work could consider nonlinear systems.
\bibliography{ifacconf}             
%
\end{document}